\documentclass{aa}   
\usepackage{graphics}
\usepackage{psfig}
\begin{document}
  \thesaurus{08         
              (03.20.2;             
               08.09.2 S140 IRS1;   
               08.06.2;             
               08.23.3)             
}
\title{Diffraction-limited bispectrum speckle interferometry and 
speckle polarimetry of the 
young bipolar outflow source S140 IRS1}

 \author{D.~Schertl\inst{1}, Y.~Balega\inst{2}, 
T.~Hannemann\inst{1}, K.-H.~Hofmann \inst{1},
Th.~Preibisch\inst{1}, G.~Weigelt\inst{1}}

\offprints{ds@mpifr-bonn.mpg.de}

 \institute{Max-Planck-Institut f\"ur Radioastronomie,
 Auf dem H\"ugel 69, D--53121 Bonn, Germany \and
Special Astrophysical Observatory, Nizhnij Arkhyz, Zelenchuk region, 
 Karachai-Cherkesia, 357147, Russia
}

   \date{Received ; accepted }

   \maketitle
\markboth{Schertl et al.:}{The massive protostar S140 IRS1}

   \begin{abstract}
We present bispectrum speckle interferometry and speckle polarimetry of 
the deeply embedded infrared bipolar outflow source S140 IRS1,
a massive protostellar object in the L1204 molecular cloud.
Using the SAO 6 m telescope, we obtained  280 mas resolution 
polarization maps of S140 IRS1 as well as a K-band image 
with diffraction-limited resolution 
$\lambda$/D of 76 mas, which is the highest angular resolution image of 
a young outflow source ever obtained in the infrared.
Our data suggest that the 
central source is marginally resolved with a FWHM diameter of
approximately 20 mas ($\sim 20$ AU).
The most remarkable feature in our image is a bright extended and very
clumpy structure
pointing away from the central source in exactly the same direction as the
blue-shifted CO outflow lobe. A centro-symmetric pattern of high polarization 
in this feature suggests that we see scattered light from the central source. 
We interprete this feature as the clumpy inner surface of a partially
evacuated cavity
in the circumstellar envelope around IRS1, which 
has been excavated by the strong outflow from IRS1.

\end{abstract}



\section{Introduction}

The formation process of massive stars is still not well understood
(cf.~Garay \& Lizano 1999 for a recent review).
The evolutionary timescales for massive stars are very short, 
as they start to burn hydrogen while still
accreting material from the surrounding protostellar cloud.
Young massive stars also affect their environment strongly by
driving very powerful winds, molecular outflows, and jets, and 
also by emitting intense UV radiation. 
An especially interesting aspect of massive star formation is the
observational result that many massive protostars drive very 
energetic outflows that are often more massive than
the central protostar (cf.~Churchwell 1997).
To reach a better understanding of the physics of massive star formation,
observations with high spatial resolution are crucial in order to
disentangle the numerous different physical processes taking place 
simultaneously.
This is the motivation for our high-resolution study of the massive
protostellar object S140 IRS1, well known for driving a massive
bipolar molecular outflow.

S140 is an HII region at the south-east edge of the L1204 dark cloud
and part of a cloud complex located at the edge
 of a prominent infrared
emission ring, known as the Cepheus ring. This ring is probably the
result of a supernova explosion and stellar winds from massive stars
in the open cluster NGC 7160, close to the center of the
ring (cf.~Kun et al.~1987.) 

About $1'$ north-east of the S140 HII region, Rouan et al.~(1977)
detected strong far-infrared emission at a position at which no objects
could be seen in visible light.
Harvey et al.~(1978) found that the spectral 
energy distribution of this infrared source, called S140 IRS,
 strongly increases between 
2~$\mu$m and 100~$\mu$m and estimated an infrared luminosity of
$\sim 2\times 10^4\,L_\odot$ (for a presumed distance of 1 kpc).
Beichman et al.~(1979) carried out 20~$\mu$m observations of this region 
and were able to
resolve the infrared emission into three individual sources. 
At~20 $\mu$m, the dominant source IRS1 is 7 -- 10 times brighter
than the two other sources IRS2 and IRS3.
The luminosity of IRS1  was estimated to be $L \approx 5 \times 10^3\,L_\odot$
(Lester et al.~1986), suggesting it to be a deeply embedded
($A_V \approx 30$ mag; Harker et al.~1997) early B-type star
with a mass of about $10\,M_\odot$. 
However, these estimates of the stellar parameters appear to 
be quite uncertain. 
A determination of the extinction towards IRS1 
based on 3~$\mu$m ice-band spectroscopy by Brooke et al.~(1996) 
yielded a column density of $N_{\rm H} = 1.4\times 10^{23}\,{\rm cm}^{-2}$
corresponding to an extinction of $A_V \sim 70$ mag. 
This would
suggest a somewhat higher luminosity and thus also a higher mass
for IRS1.

A strong molecular CO outflow in the S140 IRS region was first detected by
Blair et al.~(1978).  The total outflow mass was estimated to be
$\sim 64\,M_\odot$ (Bally \& Lada 1983), i.e.~about 6 times more
than the mass of the central protostar.
Minchin et al.~(1993) studied the CO molecular
line emission in the S140 region with the JCMT and found a bipolar
outflow morphology. 
S140 IRS1 lies just in the middle between the 
blue- and red-shifted outflow lobes; since the other two infrared sources
IRS2 and IRS3 are clearly {\em not} located on the outflow axis,
IRS1 can reliably be assumed to be the source of this outflow.
The position angle of the outflow axis is about $160\degr$.
Since the blue- and red-shifted 
outflow lobes overlap strongly, the outflow axis is believed to be
rather close to the line of sight.

In previous studies, 
near-infrared images of S140 were
obtained for example by Harker et al.~(1997) and Yao et al.~(1998).
Near-infrared polarization maps were reported by 
Lenzen (1987), Whitney et al.~(1997), and Yao et al.~(1998).
These observations, however, were seeing-limited 
and therefore  did not have the resolution required 
to study the inner environment
($\la 1000$ AU) of the central source S140 IRS1.
In order to get a better insight into the nature of this
interesting object, we have carried out bispectrum speckle interferometry
and speckle polarimetry of S140 IRS1.


\section{Observations}

The speckle interferograms were obtained with the 6~m telescope at the Special 
Astrophysical Observatory (SAO) in Russia on 19 September 1999.
The data were recorded  through a $K'$-filter with a central 
wavelength of 2.165 $\mu$m and a bandwidth of 0.328 $\mu$m.
5552 speckle interferograms of S140 IRS1 and 5736 speckle interferograms of
the reference stars HIP 110410 and HIP 110498 were taken with
our Hawaii array speckle camera.
The exposure time per frame was 150 ms, the pixel size
was 27.0\,mas and seeing was  $\sim 1''$.
The images were reconstructed using the 
bispectrum speckle interferometry method (Weigelt 1977; Lohmann et al.~1983;
Weigelt et al.~1991). The object power spectrum was determined 
with the speckle interferometry method (Labeyrie 1970). The
bispectrum of each frame consisted of 100 million elements. 
The resulting image (Fig.~1, upper right) has a diffraction-limited resolution
of 76\,mas. In order to increase the SNR we also reconstructed an image
with a reduced resolution of 130~mas (60\% of the diffraction limit),
which is shown in Fig.~1, upper left.

The speckle polarimetry observations were carried out on 13 June 1998
using the SAO 6~m telescope, 
our two-beam polarimeter, and a NICMOS 3 camera. The data recording
parameters were as follows: 
2800 polarized object speckle frames and 2800 reference star (HIP
109981) frames  recorded through 
a $2.11\,\mu$m-filter with central wavelength~/~FWHM bandwidth of
2.110~$\mu$m~/~0.192~$\mu$m, 
2400 polarized object speckle  frames and 2400 reference star (HIP
110498) frames recorded through 
a $K'$-filter with central wavelength/FWHM bandwidth of
2.165~$\mu$m~/~0.328~$\mu$m, 
exposure time per frame 150~ms,  pixel size 30.1~mas, field of view
$7.7'' \times 7.7''$
(256 $\times$ 256 pixels), and seeing $1''$. Our 2-beam speckle polarimeter
has the following standard 
design: achromatic K-band collimator objective  to obtain a parallel
beam behind the f/4 primary 
focus of the SAO 6 m telescope, rotatable $\lambda/2$ plate for
rotating the polarization direction, 
Wollaston prism to obtain two polarized images separated by $2.9''$ on the
detector, and a second 
achromatic K-band objective behind  the $\lambda/2$ plate and Wollaston
prism (more technical details of 
the speckle polarimeter will be described elsewhere). Polarized speckle 
interferograms were recorded 
at the four $\lambda /2$ plate positions $0\degr$, $22.5\degr$, $45\degr$, and $67.5\degr$.
For each of these four data 
sets and the corresponding reference star data sets, polarized images were
reconstructed using the bispectrum 
speckle interferometry method. In each of the four reconstructed images
two images of the object were 
obtained which were polarized parallel and perpendicular to the
separation vector of the two images. 
From these polarized  images polarization maps were reconstructed using
the standard technique (see e.g.~Whitney et al. 1997 and references therein). 
In order to improve the SNR of the images, the resolution 
was reduced to 280~mas. In Fig.~1 (lower left) polarization vectors are
plotted at the positions where 
the SNR of the degree of polarization (derived by splitting the data
sets) is better than 5 and the 
error of the polarization vectors is smaller than $10\degr$.

\section{Results}

\begin{figure*}
\begin{minipage}[t]{19cm}
\parbox{18.9cm}{
\parbox[c]{8.1cm}{\centerline{\psfig{figure=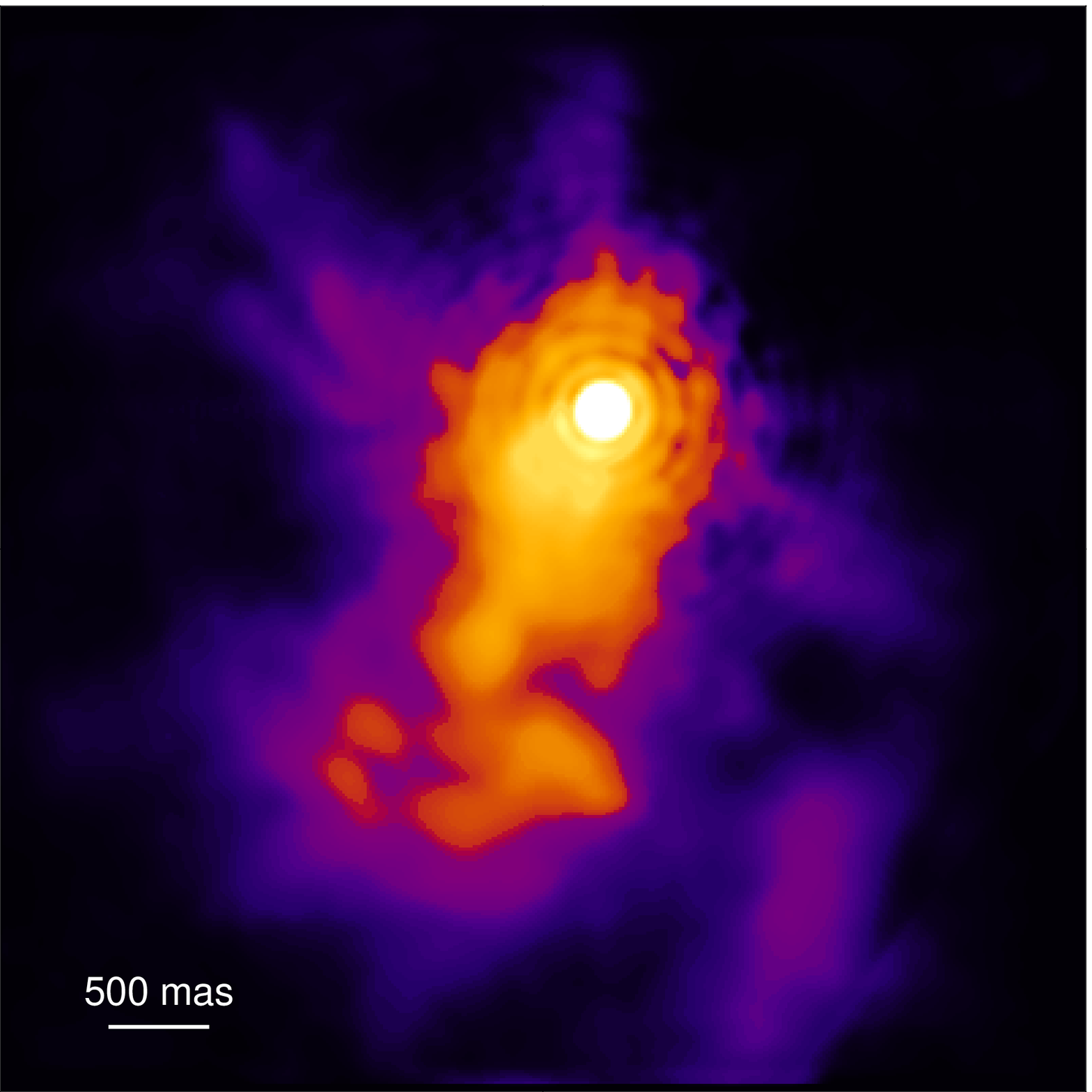,width=8.0cm}}}\hspace{2mm}
\parbox[c]{9.6cm}{\psfig{figure=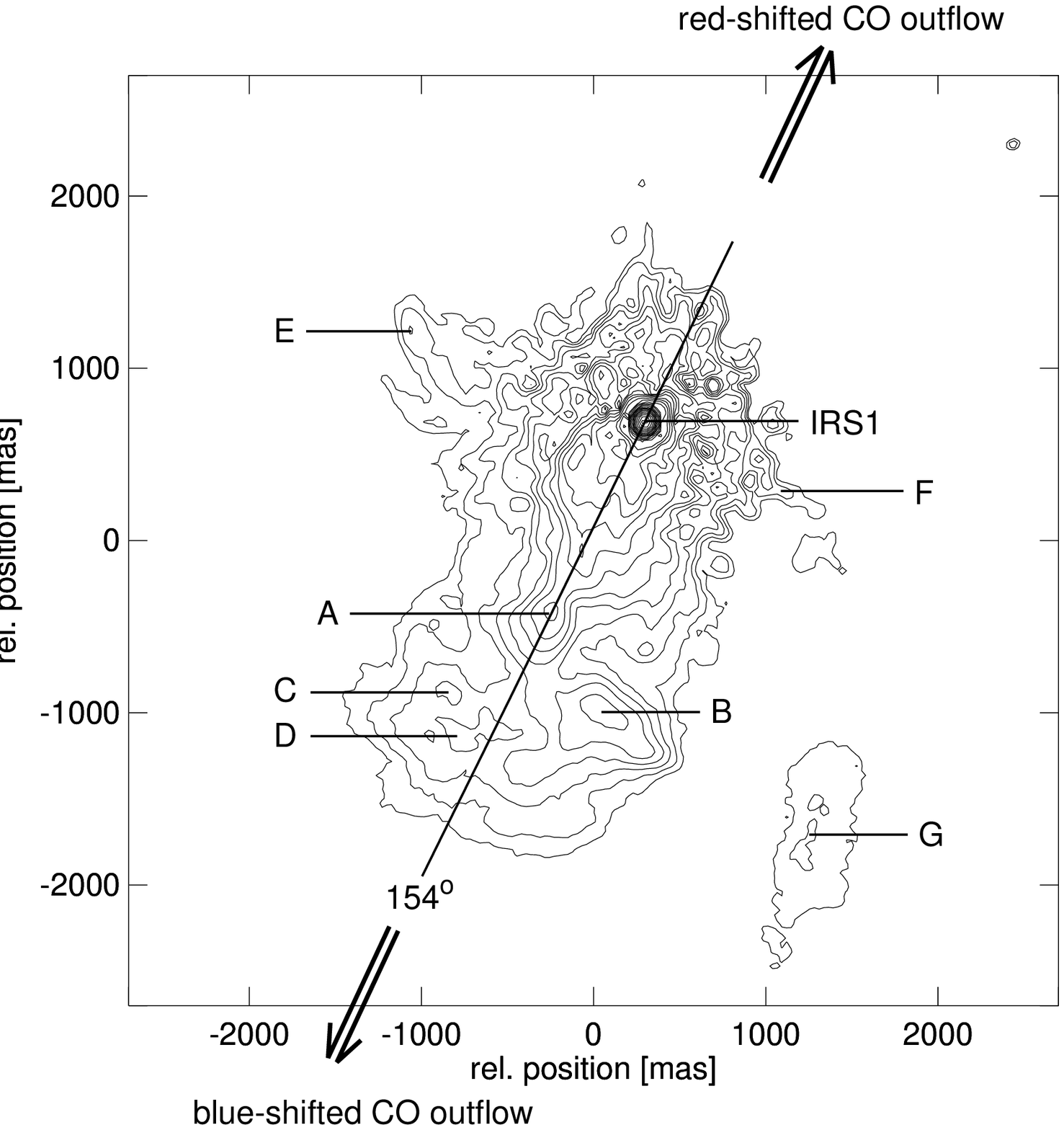,width=9.5cm}}\vspace{0mm}

\parbox{9.7cm}{\vspace{1.5cm}

\parbox{4.9cm}{\psfig{figure=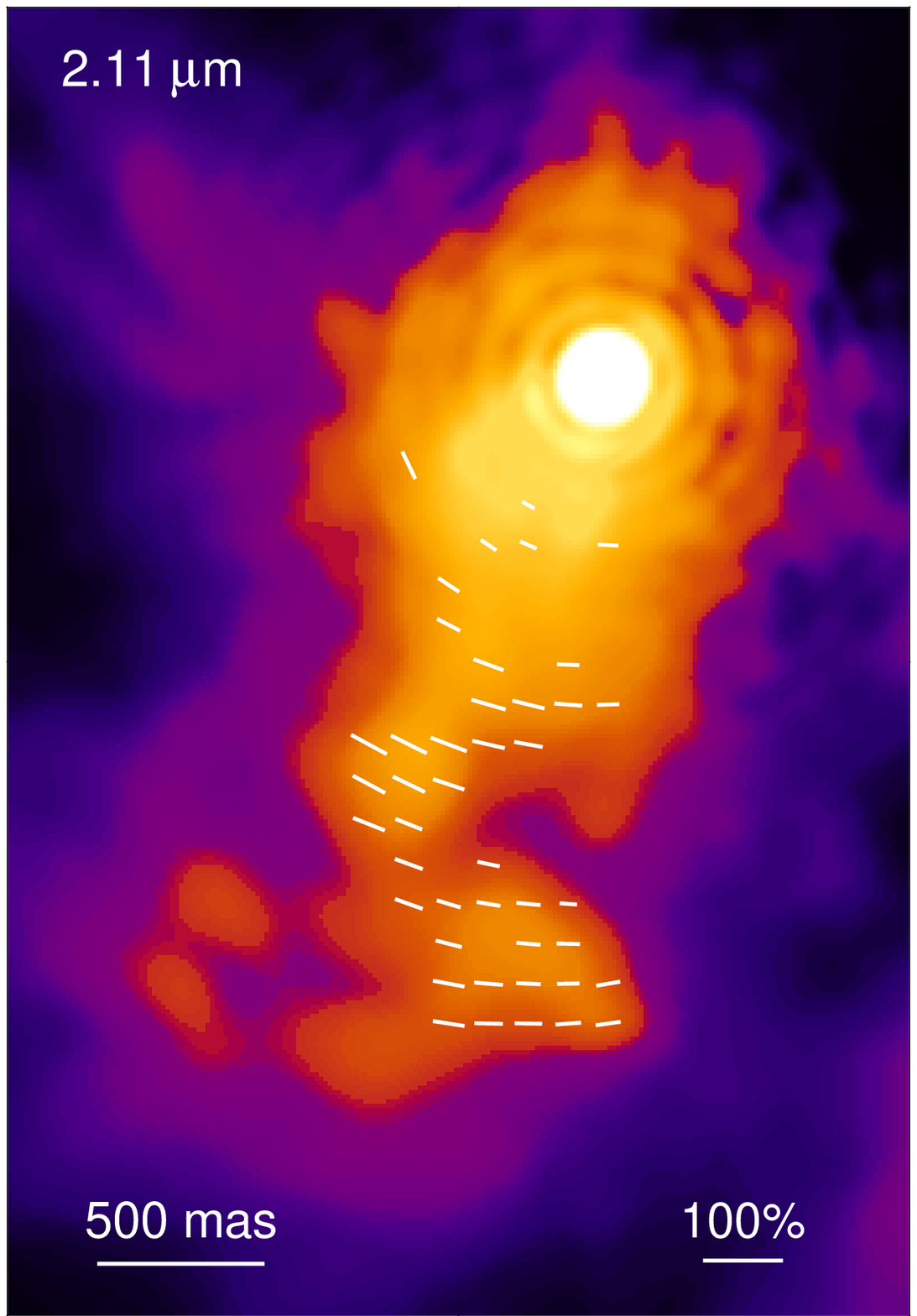,width=4.8cm}}
\parbox{4.8cm}{\psfig{figure=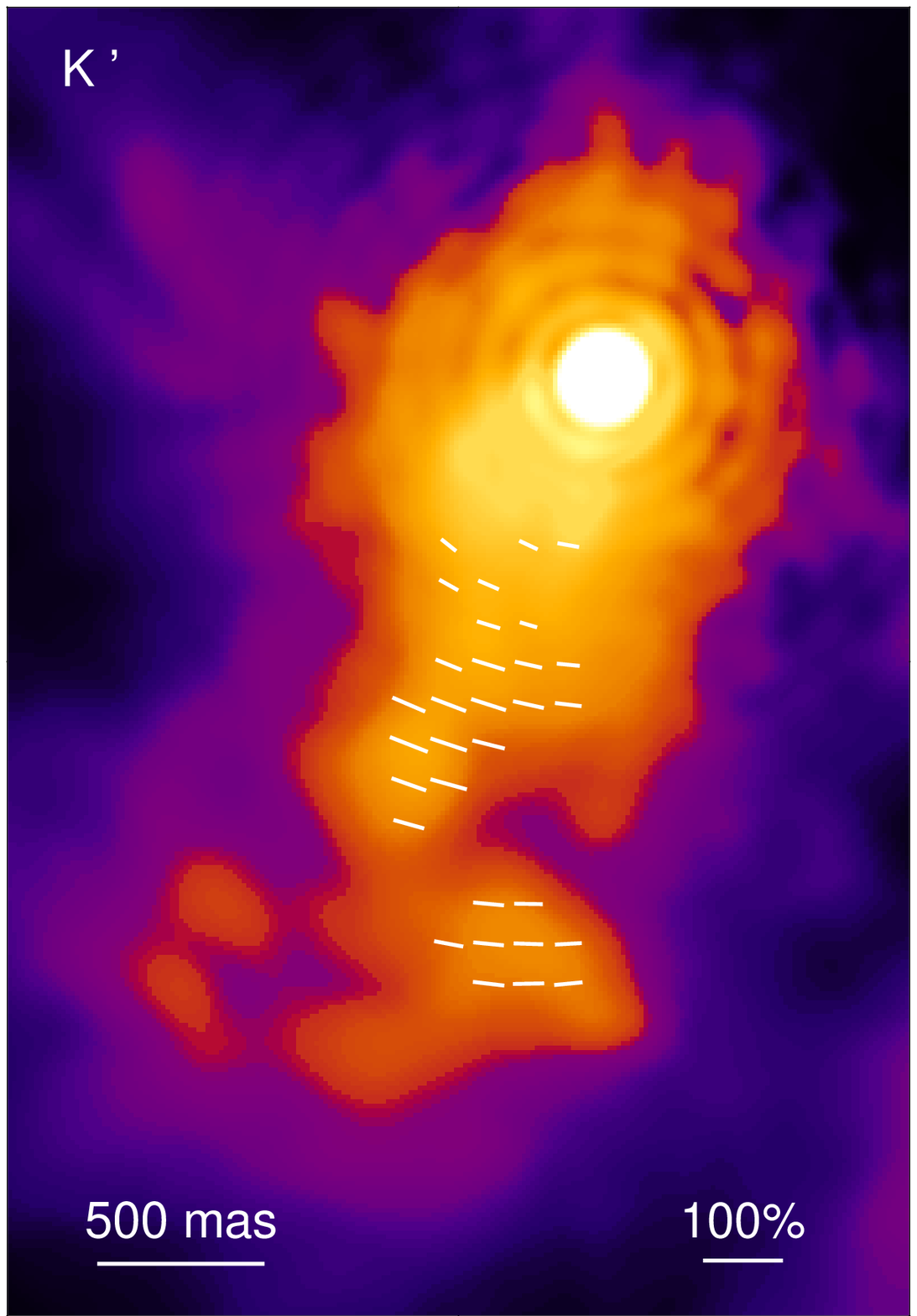,width=4.8cm}}}}
\vspace{-7.9cm}

\hspace{10.3cm} \parbox[t]{7.5cm}{
\caption{
{\bf upper left:} Color representation of our 160 mas resolution
$K$-band image of S140 IRS1 reconstructed by the 
bispectrum speckle interferometry method. Blue areas are 
$\sim 8$ mag fainter, red areas (e.g.~clumps C and D) are
$\sim 7$ mag fainter than the peak intensity.
\newline
{\bf upper right:} Contour representation of the diffraction-limited 
76 mas resolution $K$-band image with annotation;
the contour level intervals are 0.3 mag, down to 7.9\,mag\,difference relative
to the peak intensity.  The thin line marks the direction of the
symmetry axis of the
extended structure with a position angle of $154\degr$; the thick arrows
mark the directions of the red- and blue-shifted CO outflows
(position angle $155\degr$).\newline
{\bf lower left:} Color representation of the central part
of the $K$-band image with overlaid polarization vectors 
(for details see Section 2) derived 
from the $2.11 \mu{\rm m}$ data (left) and the $K'$ data (right).
 A $P = 100\%$ vector appears in the lower right corner
 for reference. 
In all images north is up and east is to the left.\label{images}
}}
\vspace{0.5cm}

\end{minipage}

\end{figure*}

\subsection{Morphology in the $K$-band image}

In Fig.~1 we present our reconstructed 
$K$-band images of S140 IRS1.
Our images show a wealth of previously unseen details.
The central source IRS1 has a point-like appearance.
A series of diffraction rings are clearly visible around IRS1.
While the first ring is perfectly closed,
the following diffraction rings are broken.
A detailed investigation of the data indicates that the 
central source is marginally resolved with a FWHM Gaussian diameter 
of approximately 20 mas (corresponding to a  physical size of $\sim 20$
AU at the assumed distance of 900 pc) and is probably elongated in a
north-south direction.

Our images show a
 bright extended feature of diffuse emission pointing from IRS1
towards the south-east. This emission has a very clumpy structure
 with several prominent knots as indicated in our contour image.
Clump A is located on the axis defined by the
general shape of the emission structure.
Three other clumps, denoted as B, C, and D, can be seen 
to the right or left of the axis.
Additionally, a rather extended feature of emission, denoted as G,
is seen south-east of IRS1.
Nevertheless, the general shape 
of the diffuse emission follows a well defined direction.
The line connecting the central maximum with clump A has a
position angle of $154\degr\pm 3\degr$ (measured counter-clock wise
from north). The two features denoted as E and F seem to trace a 
structure that is oriented roughly perpendicular to the outflow direction.

\subsection{Polarization maps}

Our polarization maps show that 
the light from the elongated feature is strongly polarized, with
degrees of polarization ranging up to 50\%. Our two polarization maps,
which were taken in slightly different filters (see above),
agree very well.
If we draw lines perpendicular to each polarization vector in our maps,
nearly all these lines cross within less than $0.3''$ of the central
intensity peak.  Thus, our maps are very consistent with a
centro-symmetric polarization pattern around the central source.
Comparison of our polarization maps with theoretical simulated 
polarization maps (e.g.~Fischer et al.~1996) suggests
that the light from the  elongated feature is scattered
light, originating from the central source.

\section{Interpretation of the observed structures}

The near-infrared emission is dominated by the central intensity peak,
that contains $\sim 35\%$ of the total flux in our $5'' \times 5''$ image.
This peak probably marks the location of the central protostar S140 IRS1.
An alternative explanation might be that the protostar is 
too deeply embedded to be directly visible in our image, and the
intensity peak might be light from the protostar that is
scattered at the walls of the outflow cavity.
However, the observed polarization pattern which is 
centro-symmetric around the intensity peak strongly suggests that the
location of the illuminating source 
actually is at the intensity peak of our image.
The observed ratio of diffuse versus direct light
together with typical optical properties of circumstellar dust
(e.g.~Preibisch et al.~1993) suggests an extinction of 
$A_K \ga 4$ mag ($A_V \ga 35$ mag) 
along the line-of-sight to the protostar, i.e., the
dominant intensity peak seems to be a highly extincted image of the
central protostar.
The possibly elongated structure of the 
central source might be caused by bright emission from the
innermost part of the larger extended structure,
or might indicate that the central object is a binary star.

The most striking feature in our image is the bright extended and very
clumpy  structure pointing from IRS1 to the south-east.
Interestingly, the position angle of this structure of $154\degr\pm 3\degr$ 
matches the direction of the blue-shifted CO outflow very well:
the $^{12}$CO  map of Minchin et al.~(1993) gives a position angle of 
 $155\degr\pm 15\degr$ for the blue-shifted outflow lobe. 
This strongly suggests that
 the elongated feature in our image is related to the blue-shifted CO outflow. 
The axis of the CO outflow is known to be closer to the line-of-sight
than to the plane of the sky (Minchin et al.~1993), 
i.e.~the south-eastern lobe of the outflow
is oriented roughly towards us, while the north-western lobe is pointing
away from us. 
This suggests the following interpretation of the observed features:

The central object IRS1 is deeply embedded
in a dense circumstellar envelope or perhaps a thick circumstellar disk. 
The outflow has cleared a cavity in the circumstellar
material and what we see as the bright extended structure south-east 
of IRS1, is light from the central protostar that is scattered by dust grains
at the inner wall of this outflow cavity into our direction.
The very clumpy appearance of the emission suggests that the
surface of the cavity wall is not smooth but highly structured. 
This might be the signature of the violent interaction between
the outflowing material and the circumstellar envelope.

Our image traces the diffuse emission out to at least about $2''$
away from the central source.  This suggests that the size 
of the circumstellar envelope or disk around IRS1 
is at least $\sim 2000$ AU.
The asymmetric general shape of our image, i.e.~the fact that we do not
see a counter lobe north-west of IRS1,  is easily explained as
a geometrical effect.
The red-shifted outflow component, pointing in the 
north-west direction, has probably
cleared a similar cavity. That cavity, however, is not visible
in our near-infrared images because it is pointing away from us
and the light predominantly escapes in a direction away from us.

\section{Conclusions}

Our high-resolution data provide new insight into  the
circumstellar environment of S140 IRS1 and enable us to study the 
structure of the envelope around the central protostar
at scales of $\sim100$ AU for the first time.
Our images reveal bright emission that most probably is reflected light
from the inner walls of a partially
evacuated cavity in the 
circumstellar material around S140 IRS1.
Due to the high spatial resolution of our data we even can see details in the
structure of the cavity walls, which appear to have a highly
inhomogeneous and clumpy surface. The fact that the orientation of the
evacuated cavity perfectly agrees with the direction of the 
molecular outflow 
strongly suggests that this cavity has been carved out by the
strong molecular outflow from  S140 IRS1.
This demonstrates the close relationship between the structure of the 
dense inner circumstellar material around the central protostar IRS1 
at scales of a few hundred AU, and the large scale molecular outflow
at scales of several 10\,000 AU.

\acknowledgements{We thank the referee for her helpful comments.}

\end{document}